\documentclass[10pt, conference, compsocconf]{IEEEtran}

\usepackage{fancyvrb}
\usepackage[pdftex,bookmarks=true,bookmarksnumbered=true,bookmarksopen=true]{hyperref}

\hypersetup{
pdfauthor = {Matt Mullins & Perry Fizzano},
pdftitle = {Treelicious: a System for Semantically Navigating Tagged Web Pages}
}

\ifCLASSINFOpdf
   \usepackage[pdftex]{graphicx}

   \DeclareGraphicsExtensions{.pdf,.jpeg,.png}
\else

\fi

\begin{document}

\title{Treelicious: a System for Semantically Navigating\\ Tagged Web Pages}

\author{\IEEEauthorblockN{
Matt Mullins}
\IEEEauthorblockA{Department of Computer Science \\
Western Washington University \\
Bellingham, WA USA \\
mtt.mllns@gmail.com}
\and
\IEEEauthorblockN{Perry Fizzano}
\IEEEauthorblockA{Department of Computer Science \\
Western Washington University \\
Bellingham, WA USA \\
perry.fizzano@wwu.edu}
}

\maketitle

\begin{abstract}Collaborative tagging has emerged as a popular and effective method for organizing and describing pages on the Web.  We present Treelicious, a system that allows hierarchical navigation of tagged web pages.  Our system enriches the navigational capabilities of standard tagging systems, which typically exploit only popularity and co-occurrence data.  We describe a prototype that leverages the Wikipedia category structure to allow a user to semantically navigate pages from the Delicious social bookmarking service.  In our system a user can perform an ordinary keyword search and browse relevant pages but is also given the ability to broaden the search to more general topics and narrow it to more specific topics.  We show that Treelicious indeed provides an intuitive framework that allows for improved and effective discovery of knowledge.
\end{abstract}


\IEEEpeerreviewmaketitle

\section{Introduction}

\label{sec:introduction}
Collaborative tagging has emerged as a popular and effective method for organizing and describing pages on the Web.  There exist many different sites in different domains that use the application of free-form keywords as a method for organizing and searching their content.  To name just a few: CiteULike for managing and discovering scholarly references, LibraryThing for cataloging and sharing literature, Etsy for buying and selling handmade items, and Delicious\footnote{http://delicious.com/} for organizing and sharing bookmarks.  Tagging becomes especially useful to describe non-text media like photos on Flickr and videos on YouTube.  These sites have embraced tagging as an effective and low-cost way of describing and organizing their content.  On Delicious, one of the most popular social bookmarking sites, users annotate pages with tags, usually for the selfish reason of personal organization.  Yet when this is done by many individuals, collectively rich and accurate descriptions of what these resources mean to humans materializes.  Even though users are using tags primarily to help themselves retrieve the page later, 62\% of the tags in Delicious end up identifying descriptive facts about the web resource---tags useful beyond personal organization~\cite{al-khalifa:patterns}.  This user-generated classification structure has come to be known as a ``folksonomy\footnote{http://vanderwal.net/folksonomy.html}''.

\begin{figure}[h]
\centering
\setlength\fboxsep{0pt}
\setlength\fboxrule{0.5pt}
\fbox{
\includegraphics[width=1.7in]{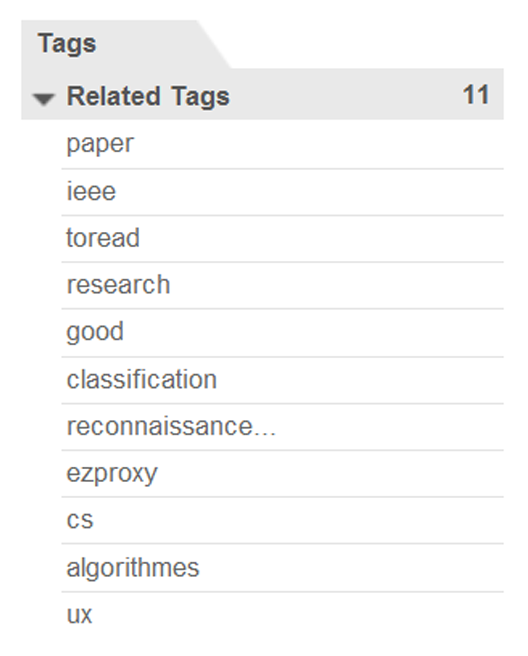}
}
\caption{A Delicious search for ``acm''' yields these ``related'' tags.  Their relation is based solely on co-occurrence information.}
\label{fig:acmdelicious}
\end{figure}

Yet these folksonomies are lacking in several ways.  First, they're flat.  There is no explicit hierarchy, synonymy, or relation information present---only simple co-occurrence data.  Second, they're ambiguous.  This is the classic problem of using words with multiple meanings and no explicit disambiguation information.  Given this lack of semantics there are only a handful of ways we can present sets of tags to the user.  A common method is to use a tag ``cloud'' with more popular tags in the cloud indicated by a larger font size.  Another method is to start with a search tag and present related tags based on which tags the search tag co-occurs with in tagged content.  This co-occurrence data can also be used to group related tags using clustering techniques as is done in Flickr.  Though all of these methods are helpful in some way, ultimately, they fail to show the semantic relationships among tags~\cite{laniado:wordnet}.  As a result, it is hard for a user to put their search into perspective.  Figure~\ref{fig:acmdelicious} shows an example of the related tags produced from a search for ``acm'' on Delicious.

This lack of structure resulting from the use of free-form tags is not encountered with more classic systems of classification like hierarchical taxonomies and library classifications.  The categories in these systems are well-defined and placed in a strict hierarchy.  Each subcategory can have only one parent category of which it is a member.  Such a structure results in clear semantic ``broader than'' and ``narrower than'' relationships among concepts.  But the strictness inherent in these classic systems presents disadvantages.  They require expert catalogers, authoritative sources of judgment, and users educated about the categories~\cite{shirky:ontology}.  It also takes work to keep them from becoming outdated as new categories are formed and old ones are restructured (e.g. ``the Soviet Union'' being reclassified as a ``Former country'').  Commenting on the restriction that each class have only one parent, Voss~\cite{voss:wikipedia} observes that ``Hierarchy seems to have a strict semantic that does not fit to the vagueness of the world.  In practice there are always several ways to classify an object  \ldots  If one uses polyhierarchy like in a thesaurus, the system is much more flexible.''

\begin{figure}[h]
\centering
\setlength\fboxsep{0pt}
\setlength\fboxrule{0.5pt}
\fbox{
\includegraphics[width=3in]{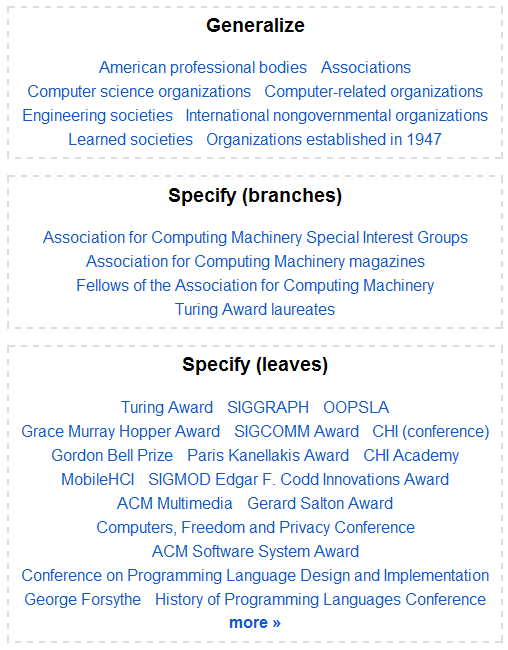}
}
\caption{A Treelicious search for ``acm'' yields these semantically related tags.  We expand on the details of Treelicious in Section~\ref{sec:methods}.}
\label{fig:acmtreelicious}
\end{figure}

And thus we come to Treelicious, the combination of a free-tagging system and polyhierarchy---the best of both worlds.  Treelicious takes the freedom and fluidity of tagging systems and leverages a thesaurus to impose semantic structure on the tags.  Navigation around the tag space now becomes more intuitive and informative since we can generalize to broader content and specify to narrower content.  As an example, contrast the ``acm'' search on Delicious in Figure~\ref{fig:acmdelicious} with our system in Figure~\ref{fig:acmtreelicious}.  Notice that the tags produced in our system are \textit{semantically} related to the current tag.  The type of relation is expressed in the grouping of the tags.

\section{Related Work}
\label{sec:related}

Since hierarchical organization aids human cognition, it is natural that researchers have explored the idea of hierarchy in folksonomy.

Heymann and Garcia-Molina \cite{heymann:hierarchy} convert a large corpus of tagged objects into a hierarchical structure of tags using purely statistical techniques.  They achieve this by leveraging concepts of generality and similarity users implicitly embed in their annotations and applying graph centrality algorithms to build a tree of tags.  Though the tree generated is surprisingly accurate in some places, it breaks down to a simple similarity graph in others and isn't semantically sound enough for reliable hierarchical navigation.

Instead of generating a hierarchy from the tags themselves, Laniado et al. \cite{laniado:wordnet} impose hierarchy on a set of tags using the WordNet\footnote{http://wordnet.princeton.edu/} lexical database.  In their system, when a tag is used to perform a search on Delicious, they gather a sample of tags that have been applied to each of the result pages into one big set of tags.  They then pipe these tags through a module that utilizes hypernym and hyponym hierarchy information present in WordNet to build a semantic tree of tags related to the search tag.  They also prune out WordNet nodes that don't appear in the tag set to compress their tree.  Because of this, their hierarchy is bounded by the search tag.  Though their results are nicely hierarchical, they lack a sense of completeness having only been seeded by the co-occurring tags in the local results.  There are also problems with the mapping from Delicious tags to WordNet words due to the difference in the level of formality of language (e.g. nyc versus New York City) and the prevalence of recently introduced terminology in Delicious (e.g. AJAX, Obama, Harry Potter).

Faviki\footnote{http://www.faviki.com/} is a recent social bookmarking tool that uses the newly developed Common Tag\footnote{http://www.commontag.org/} format for tags.  Mili\v{c}i\'c \cite{milicic:faviki} points out the disadvantage of tags: they do not provide information about their meaning.  When there are multiple meanings for one word or multiple words for one meaning traditional tagging systems fail to distinguish.  This leads to inappropriate connections and hindered search and browsing.  Instead of tagging bookmarks with free-form text as is traditional, Faviki uses Wikipedia\footnote{http://www.wikipedia.org/} concepts as tags.  These so-called ``semantic tags'' are unique, well-defined, and part of the controlled vocabulary of Wikipedia.  This connection to Wikipedia enables tying in to a rich network of semantics, but it forces a vocabulary on the users, which is problematic.  To drive home this point consider the task an author faces when choosing applicable ``Categories and Subject Descriptors'' from the ACM Computing Classification System and choosing their own ``Keywords'' when classifying their research paper.  Clearly choosing keywords is less of a burden than trying to select a designation from a strict hierarchy.  Butterfield \cite{butterfield:aside} sums it up well: ``[To a user] Free typing loose associations is just a lot easier than making a decision about the degree of match to a pre-defined category (especially hierarchical ones).  It's like 90\% of the value of a `proper' taxonomy but 10 times simpler.''

Treelicious is a natural extension from these ideas.  Our prototype uses Delicious as its tagging system and leverages the Wikipedia category structure to semantically navigate tagged web pages.  By utilizing Wikipedia we have the advantage of a dataset with more than 2.9 million (English language) concepts and 470 thousand categories as opposed to the 150 thousand words in Wordnet.  Wikipedia is also collaboratively edited by thousands of editors and so adapts well to change.  It also contains redirects from popular misspellings, acronyms, and alternative terminology, and thus is more forgiving of the diverse and sometimes informal language of tags found on Delicious.  Instead of grabbing a set of pages and then generating a partial hierarchy from only the related tags, we place the user in the context of a complete concept hierarchy, but limit them to actions that move them through the hierarchy one step at a time.  We stress that this hierarchy is complete; we don't prune out concepts that don't have any results in Delicious because we feel like these nodes act as valuable stepping stones for meaningful navigation and offer knowledge just by being in the hierarchy.  We also stick to the practice of using free-text tags instead of relying on Common Tags because it is more prevalent in current tagging systems and is clearly easier.

In the following sections we detail and demonstrate our system.

\begin{figure*}
\centering
\includegraphics[width=7in]{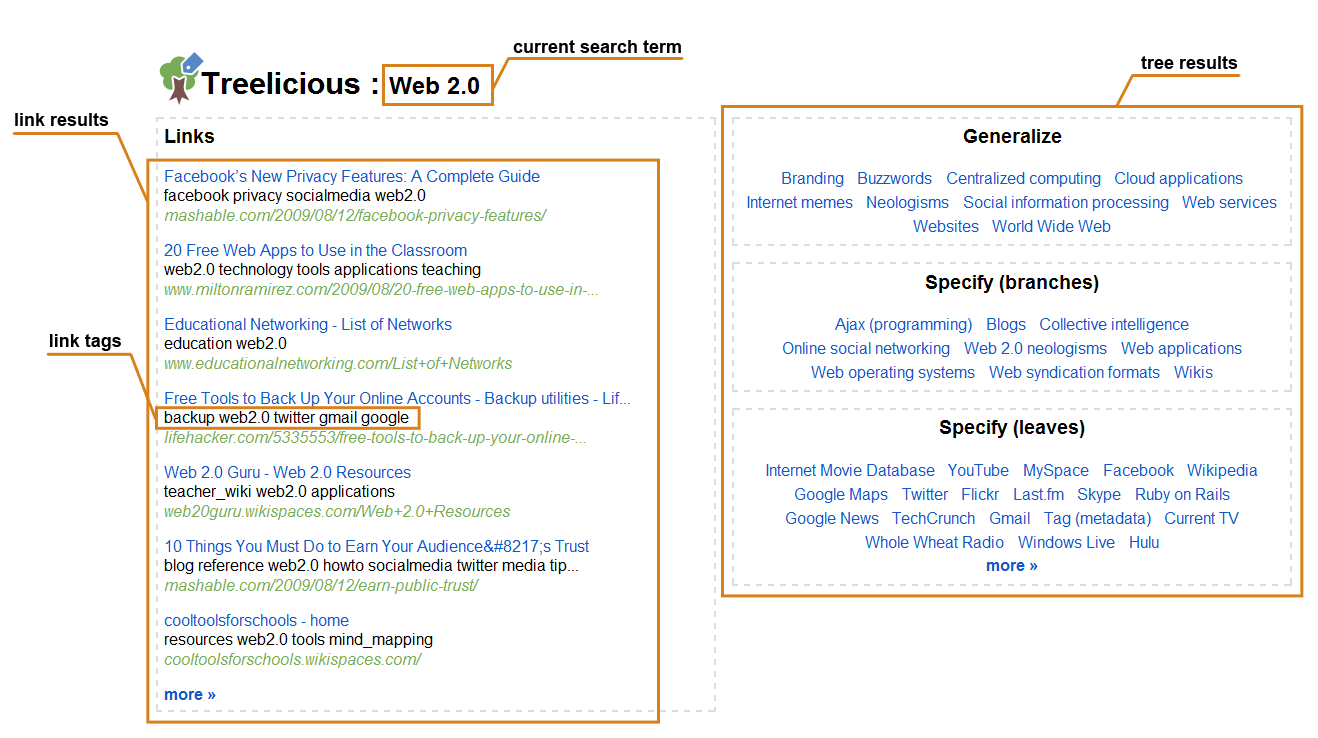}
\caption{The different components of the results page for the search term ``Web 2.0''.  Broader terms fall under the ``Generalize'' box.  Narrower terms are split up into ``Specify (branches)'', which offer further room for refinement (having narrower terms themselves), and ``Specify (leaves)'' which are terminal terms that do not allow further specification.}
\label{fig:web2}
\end{figure*}

\section{Terminology}
\label{sec:terminology}

Before we dive into the specifics of our datasets and methods, we first detail the terminology we use by walking through an example search.

A user first supplies a search keyword that is disambiguated into the \textit{current search term}.   Treelicious then returns a \textit{results page} for the term.  The results page is composed of two main sections; the \textit{link results} and the \textit{tree results}.  The link results provide links to web pages that are bookmarked in Delicious.  The tree results give a glimpse of the immediate hierarchy in which the current search term resides.

See Figure \ref{fig:web2} for more detail.


\section{Datasets}
\label{sec:datasets}
To access the data present in Wikipedia we use DBpedia~\footnote{http://dbpedia.org/}, a community effort to extract structured information from Wikipedia.  They represent this data using RDF (Resource Description Framework) and provide an online SPARQL (SPARQL Protocol and RDF Query Language) endpoint for querying this data.    

To access Delicious we use the JSON (JavaScript Object Notation) data feeds they provide.  For our purposes we're interested in finding bookmarks by tag and Delicious allows us to access a limited number of ``popular'' bookmarks and ``recent'' bookmarks associated with a search tag.  Unfortunately, most tags don't have any ``popular'' results (the undocumented measure of ``popular'' is not just the most bookmarked pages), so we fall back on the ``recent'' feed to provide at least some results to the user.  Right now we simply grab the results from both feeds on demand.  This provides us only a very basic view of the page space though, as we are given only a few descriptive tags for each result.  In Section~\ref{sec:future}, Future Work, we explore alternatives to get around this limitation.


\section{Methods}
\label{sec:methods}
We begin by giving an outline of the steps our system goes through to generate a user's first results page.

\begin{enumerate}
\item User enters keyword to search Treelicious
\item Keyword disambiguated
\item Map from disambiguated term to DBpedia concepts
\item Generate tree results
\item Fetch link results
\end{enumerate}

Once a user has performed the initial entry into the system, they are given the opportunity to navigate to broader and narrower concepts, in which case the system repeats steps 3-5.  These steps are explained in greater detail in the subsections below.

\subsection{Search and Disambiguation}
To enter the system initially we need some method for arriving at an identifier that actually exists in DBpedia.  Though there are a few different ways we considered approaching this, we ended up using DBpedia's own lookup API\footnote{http://lookup.dbpedia.org/api/search.asmx}.  When a user enters our site they can specify a search keyword.  The system passes this keyword to the API and retrieves a list of possible ``disambiguation'' terms with short descriptions (taken from the description of the corresponding Wikipedia article).  The user can then choose which concept they meant and the system will direct them on to the appropriate results page.

\subsection{Mapping}

Wikipedia has a few different kinds of pages in which we are interested.  Namely, \textit{Article} pages, \textit{Category} pages, \textit{Redirect} pages, and \textit{Disambiguation} pages.  

Article pages are the pages that provide the encyclopedic information and are the ones with which most people are familiar when it comes to Wikipedia.  Category pages list the Articles and Categories that are members of a particular concept.  Redirect pages contain no articles themselves but direct users to an Article from an alternative term for the article.  For example the Redirect page for ``nyc'' sends users to the Article page for ``New York City''.  Finally, Disambiguation pages are used when a single term is associated with more than one article.  They list the possible candidate articles with short summary text.

The category system works like so:  each Article page is ``annotated'' with a number of categories to place the article in those categories.  These categories can be topic categories, which contain pages on a particular topic, or list categories, which contain pages whose subjects are members of a set.  There are then Category pages (denoted with a ``Category:'' prefix before the category label) that list all the pages in the category.  These Category pages are annotated with categories themselves to make the category a subcategory of others.  Thus this categorization forms a semantic tree of categories, with a root category, subcategories, and article leaves.  Wikipedia's category structure doesn't actually form a strict tree.  Treelicious is a misnomer; though the hierarchy is tree-like in that it provides upward movement to more general concepts and downward movement to more specific concepts, concepts can have more than one immediate parent.  The structure, then, is more like a directed acyclic graph, but DAGlicious didn't have quite the same ring to it. 

Often categories on Wikipedia will also have a main article that shares the same or similar name and that contains the encyclopedic content.  For example there is a ``Ford Motor Company'' article and a ``Category:Ford'' category.  These articles should be a member of their eponymous category themselves, and are denoted as the key articles for the category with special templates and sort keys\footnote{http://en.wikipedia.org/wiki/Wikipedia:Categorization}.  However DBpedia does not (yet) extract this information so all we have are the facts that the article is a member of the category and that they share the same or similar names.  Often this is enough information to associate the two pages in the system, but there are some cases where our simplistic technique fails.  

This association is important to us because Treelicious combines the two resources when it can to provide both broader and narrower concepts on one page.  To assist this matching between article and category the system utilizes redirect information present in DBpedia.  For example, the article ``Food'', the redirect ``Foods'' (which points to the article ``Food''), and the category ``Category:Foods'' are all utilized to form one page and allow for smoother navigation throughout the hierarchy.

\subsection{Tree Results}
After the mapping has been performed the system grabs the terms that make up the tree.  It first queries for a set of broader terms as the first step in creating the ``generalize'' results.  The system then queries for a set of narrower terms.  Because categories in Wikipedia can have both sub-articles and subcategories, we will have two different sets of ``specify'' terms.  While the system could combine these two sets, we found that it is helpful to leave them separate and call the terms that are pulled from subcategories ``branches'' and the terms that are pulled from sub-articles ``leaves''.  This way a user can glance at the results and know which terms cannot become more specific and which offer more opportunity for specification.  

Furthermore, we refine the presentation of the leaf terms (which correspond to sub-articles) because there is page link information present in DBpedia for them.  By aggregating in the query how many other Wikipedia articles contain links to the result article, the system orders our list of terms by how connected they are with the rest of Wikipedia.  The effect of this effort can be seen in Figure \ref{fig:web2} where the top 20 leaf terms (all very relevant) are shown out of nearly 1000 possible results (many of which are not very relevant).

\subsection{Link Results}
To give the user a set of link results we query Delicious.  The user is then presented with a sample of the results favoring the popular ones.  The interface provides the title of the link which doubles as a hyperlink to it, a brief list of link tags (top tags in the case of popular and co-occurring tags in the case of recent) to summarize the link, and a preview of the link URL.

We note that the links returned by our system are confined to those tagged in Delicious and more specifically to the ``popular'' and ``recent'' pages which we have access to via the limited Delicious data feeds.  We expand on the possibilities for this feature in Section~\ref{sec:future}, Future Work.


\section{Analysis}
\label{sec:analysis}
 In the future we plan to carefully construct an array of user studies to quantify the usefulness of our system in terms of what types of knowledge discovery tasks are best performed by a system like Treelicious when compared to current tools.  For now we present an illustrative example and contrast the results obtained in Treelicious with current technologies.  

In computer science we are often confronted with the task of deciphering new jargon and determining if the jargon is just a new name for an old concept or if it is truly something new.  Learning about new jargon requires us to see what terms it is related to and how it fits in with our already vast knowledge of concepts in computer science as a whole.  

For instance, suppose we keep seeing the term ``ruby on rails'' thrown around but we have no experience with it ourselves.  We could use Treelicious to get a quick sense of what it is and why we should care.  A user examining the results page for this search would learn that Ruby on Rails is a leaf topic (very specific relative to the rest of the tree).  However, its generalizations reveal that it is a web application framework programmed in Ruby and is a member of this other concept we hear thrown around, ``web 2.0''.  If the user was happy here, he could follow any of the many web page results for tutorials and guides provided in the link results.  If, though, he was curious about other frameworks and technologies grouped under ``web 2.0'' he could follow that branch.  Refer back to Figure~\ref{fig:web2}.  

A popular interest on Delicious, the Web 2.0 tag gives the user many pages to explore.  Some of the generalizations may confirm beliefs the user already had; that it's a buzzword and a neologism.  The specify branches show some overarching terms that comprise the field: Ajax programming, blogs, wikis, online social networking, and web applications.  If the user were to follow these links he'd have further opportunity to narrow his exploration.  The specify leaves give many instances of technologies, ideas, people, companies, applications, and services that are members of the Web 2.0 field.

Another useful knowledge acquisition task that can be carried out in Treelicious is to sidestep from a search term to see sister concepts.  For instance, by generalizing to ``Web application frameworks'' and then observing the ``Specify (leaves)'' box, we see that Ruby on Rails is similar to ASP.NET and Drupal.  In some sense Treelicious supports learning by example with this sidestep functionality.

From our Ruby on Rails example we can see some of the opportunities for knowledge discovery Treelicious gives us that traditional systems do not.  We now contrast our system with Delicious alone, Wikipedia alone, and Web search.  

\begin{trivlist}
\item \textbf{Delicious search}

Compared to a straight Delicious search where we are given a simple list of co-occurring tags, Treelicious displays our search term in the context of a tree of semantic relationships.  When a user wants to refine her search, the semantic hierarchy provided by Treelicious offers richer options for navigation.  It has been shown time and again that categorization and hierarchy can facilitate search and retrieval due to the fact that categorization is a fundamental process in human cognition.  

Further, the synonymy information pulled from DBpedia allows Treelicious to equate multiple terms with the same meaning.

\item \textbf{Wikipedia search}

Compared to a straight Wikipedia search, Treelicious has the advantage of a more intuitive interface to generalization and specification. It also has a source of richly tagged links deemed important by Delicious users and links to new pages whose content has not yet been (and may never be) incorporated into Wikipedia.  These link results offer both a more in-depth view of a topic through primary sources and also a more current view of new developments.  This is a large advantage when searching for topics that aren't well covered on Wikipedia.  

\item \textbf{Web search}

Compared to a straight Web search, Treelicious has the advantage of links that have been deemed relevant by humans instead of an automated scheme based on the link structure of the Web.  We also have a rich tag space for each of our links.  These tags provide a description of the page that is more representative of what the page actually means to humans than a simple scrape of the words in the document (30\% of the tags used to annotate the page have been shown to be relevant but do not appear in the page text \cite{heymann:search}).

\end{trivlist}


\section{Future Work}
\label{sec:future}
There are a number of elements of Treelicious that could be improved, the main one being disambiguation of link results.  The main barrier to this, as hinted to in Section~\ref{sec:datasets}, is our limited access to Delicious bookmarks.  Ideally we would retrieve a list of the most bookmarked pages tagged with a certain tag, along with all the tags applied to each page.  As it is now, the links in the ``popular'' feed are ranked by an undocumented measure involving multiple variables including recentness, and some tags don't even have any popular results.  If we had access to a more dense tag space for each link we could theoretically use our position in the term hierarchy along with the tag ancestry (the list of continually broader terms in the hierarchy) to filter to link results that are only about this particular meaning of the tag.  Tesconi et al. \cite{tesconi:semantify} tackle a similar problem.  This would hinge on the data containing natural hierarchical relations, and indeed Golder and Huberman \cite{golder:structure} observe this.  Heymann et al. summarize their observation well by saying this is ``a general feature of tagging data both because users appear to tag from their own personal mental taxonomies (leading to multiple levels on a per-user basis) and because different users have different context-specific basic levels (the level of detail at which a user views an object).''~\cite{heymann:hierarchy}


\section{Conclusion}
\label{sec:conclusion}
We have proposed and prototyped a system for semantically navigating tagged web pages.  By leveraging the Wikipedia collaborative encyclopedia on top of the Delicious social bookmarking folksonomy we demonstrate a prototype that uses the largest, most up to date, and most collaborative information on the Web.  We've shown that Treelicious presents an intuitive way to navigate and explore this rich data in an effective and concise manner that fills a unique void not occupied by web search results, Delicious results, or Wikipedia results alone.

\bibliographystyle{IEEEtran}
\bibliography{IEEEabrv,treelicious}

\begin{thebibliography}{10}
\providecommand{\url}[1]{#1}
\csname url@samestyle\endcsname
\providecommand{\newblock}{\relax}
\providecommand{\bibinfo}[2]{#2}
\providecommand{\BIBentrySTDinterwordspacing}{\spaceskip=0pt\relax}
\providecommand{\BIBentryALTinterwordstretchfactor}{4}
\providecommand{\BIBentryALTinterwordspacing}{\spaceskip=\fontdimen2\font plus
\BIBentryALTinterwordstretchfactor\fontdimen3\font minus
  \fontdimen4\font\relax}
\providecommand{\BIBforeignlanguage}[2]{{%
\expandafter\ifx\csname l@#1\endcsname\relax
\typeout{** WARNING: IEEEtran.bst: No hyphenation pattern has been}%
\typeout{** loaded for the language `#1'. Using the pattern for}%
\typeout{** the default language instead.}%
\else
\language=\csname l@#1\endcsname
\fi
#2}}
\providecommand{\BIBdecl}{\relax}
\BIBdecl

\bibitem{al-khalifa:patterns}
H.~S. Al-Khalifa and H.~C. Davis, ``Towards better understanding of folksonomic
  patterns,'' in \emph{HT '07: Proceedings of the Eighteenth Conference on
  Hypertext and Hypermedia}.\hskip 1em plus 0.5em minus 0.4em\relax New York,
  NY, USA: ACM, 2007, pp. 163--166.

\bibitem{laniado:wordnet}
D.~Laniado, D.~Eynard, and M.~Colombetti, ``Using wordnet to turn a folksonomy
  into a hierarchy of concepts,'' in \emph{Semantic Web Application and
  Perspectives - Fourth Italian Semantic Web Workshop}, Dec. 2007, pp.
  192--201.

\bibitem{shirky:ontology}
\BIBentryALTinterwordspacing
C.~Shirky. (2005) Ontology is overrated: Categories, links, and tags. [Online].
  Available: \url{{http://www.shirky.com/writings/ontology\_overrated.html}}
\BIBentrySTDinterwordspacing

\bibitem{voss:wikipedia}
\BIBentryALTinterwordspacing
J.~Voss, ``Collaborative thesaurus tagging the wikipedia way,'' \emph{CoRR},
  2006. [Online]. Available: \url{{http://arxiv.org/abs/cs/0604036v2}}
\BIBentrySTDinterwordspacing

\bibitem{heymann:hierarchy}
\BIBentryALTinterwordspacing
P.~Heymann and H.~Garcia-Molina, ``Collaborative creation of communal
  hierarchical taxonomies in social tagging systems,'' Stanford InfoLab, Tech.
  Rep. 2006-10, April 2006. [Online]. Available:
  \url{http://ilpubs.stanford.edu:8090/775/}
\BIBentrySTDinterwordspacing

\bibitem{milicic:faviki}
\BIBentryALTinterwordspacing
V.~Mili\v{c}i\'c. (2008) {W3C} semantic web case study: Semantic tags.
  [Online]. Available:
  \url{{http://www.w3.org/2001/sw/sweo/public/UseCases/Faviki/}}
\BIBentrySTDinterwordspacing

\bibitem{butterfield:aside}
\BIBentryALTinterwordspacing
S.~Butterfield. (2004). [Online]. Available:
  \url{http://www.sylloge.com/personal/2004/08/folksonomy-social-classificatio%
n-great.html}
\BIBentrySTDinterwordspacing

\bibitem{heymann:search}
P.~Heymann, G.~Koutrika, and H.~Garcia-Molina, ``Can social bookmarking improve
  web search?'' in \emph{First ACM International Conference on Web Search and
  Data Mining (WSDM'08)}, February 2008, pp. 195--206.

\bibitem{tesconi:semantify}
M.~Tesconi, F.~Ronzano, A.~Marchetti, and S.~Minutoli, ``Semantify del.icio.us:
  Automatically turn your tags into senses,'' in \emph{Proceedings of the First
  Social Data on the Web Workshop {(SDoW2008)}}, 2008.

\bibitem{golder:structure}
\BIBentryALTinterwordspacing
S.~A. Golder and B.~A. Huberman, ``The structure of collaborative tagging
  systems,'' \emph{CoRR}, 2005. [Online]. Available:
  \url{{http://arxiv.org/abs/cs/0508082}}
\BIBentrySTDinterwordspacing

\end{thebibliography}

\end{document}